\documentclass[reprint,twocolumn,amsmath,showpacs,superscriptaddress,amssymb]{revtex4-1}

\usepackage{graphicx}
\usepackage{graphics}
\usepackage{amsmath}
\usepackage{dcolumn}
\usepackage{bm}

\begin{document}


\title{ Strain-controlled Insulator-Metal Transition in YTiO$_3$/SrTiO$_3$ Superlattices: A First-Principles Study}
\author{Xue-Jing Zhang}
\affiliation{Beijing National Laboratory for Condensed Matter Physics, Institute of Physics, Chinese Academy of Sciences, Beijing 100190, China}
\affiliation{School of Physical Sciences, University of Chinese Academy of Sciences, Beijing 100190, China}
\author{Peng Chen}
\affiliation{Beijing National Laboratory for Condensed Matter Physics, Institute of Physics, Chinese Academy of Sciences, Beijing 100190, China}
\affiliation{School of Physical Sciences, University of Chinese Academy of Sciences, Beijing 100190, China}
\author{Bang-Gui Liu}\email{bgliu@iphy.ac.cn}
\affiliation{Beijing National Laboratory for Condensed Matter Physics, Institute of Physics, Chinese Academy of Sciences, Beijing 100190, China}
\affiliation{School of Physical Sciences, University of Chinese Academy of Sciences, Beijing 100190, China}

\date{\today}

\begin{abstract}
The structural, magnetic, and electronic properties of (STO)$_4$/(YTO)$_2$ superlattice consisting of Mott insulator YTiO$_3$ (YTO) and band insulator SrTiO$_3$ (STO) under strain are investigated by the density-functional-theory plus \emph{U} method. It is found that an insulator-metal transition occurs when a compressive strain of 0.2\% is applied. The structural analyses reveal that the presence of metallic state in such superlattices accompanies structural phase transition with restoring of inversion symmetry. Further study shows that this strain-induced structural transition makes the $d$ energy level of the interfacial Ti atoms of the YTO layer move upward due to the decreasing of the TiO$_{6}$ octahedral volume and induces the electron reconstruction in the whole superlattice systems. In addition, when the on-site interaction $U$ is changed from 5 to 4 eV, a similar insulator-metal transition also occurs in such superlattices due to the weakened electron correlation. These findings can improve our understanding of the insulator-metal transitions in such oxide superlattices.
\end{abstract}

\maketitle


\section{INTRODUCTION}

The discovery of conducting behavior at the interface between LaAlO$_3$ (LAO) and SrTiO$_3$ (STO) (both are band insulators) initiated a broad search to elucidate the origin of the interfacial metallic state.\cite{1,2,3,4} The widely cited mechanisms used to explain the conductivity at the LAO/STO interface are the polar catastrophe\cite{5,6}, oxygen vacancies\cite{7}, etc.. This discovery has also triggered huge interest in designing other STO-based heterostructures,\cite{8,9,10,11} especially heterointerfaces between Mott insulator $R$TiO$_3$ ($R$ is a trivalent rare earth ion) and STO, because the Mott insulator have a strong electron correlation and symmetrical interfaces can be formed in $R$TiO$_3$/STO systems, due to both $R$TiO$_3$ and STO have the TiO$_2$ atomic layer.\cite{12,13,14} Metal-insulator transitions (MIT) have been observed in some heterostructures, such as LaTiO$_3$/STO and GdTiO$_3$/STO, and polar catastrophe, substrate-induced deformation etc. are taken into consideration when interpreting the conductivity.\cite{15,16,17,18,19,20,21,22} In addition, the recent observation of MIT in the $d^1$ and $d^2$ perovskites driven by strain-induced changes in crystal-field splitting and in the hopping parameters has received renewed interest.\cite{23,24,25,26,27} As a prototypical Mott insulator with $d^1$, YTiO$_3$ (YTO) crystallizes in the orthorhombic distorted GdFeO$_3$-type structure characterized by $a^-a^-b^+$-type octahedral tilts in Glazer notation\cite{28} (space group \emph{Pbnm}), having experimental lattice constants of $a=5.32$ \AA, $b=5.68$ \AA, and $c=7.61$ \AA.\cite{29} Bulk YTO has been reported to behave as Mott ferromagnetic insulator with a saturation moment of 0.84 $\mu_B$/Ti$^{3+}$ and a Curie temperature $T_C$ of 30 K.\cite{30}

Here, we construct short-period YTO/STO superlattices, and study their structural, magnetic, and electronic properties, with strain applied. Our calculated results and analyses show that there exists a structural phase transition, restoring the inversion symmetry, when a compressive strain is more than 0.2\% or the on-site interaction \emph{U} is changed from 5 to 4 eV. This structure phase change induces insulator-metal transition by electronic reconstruction. More detailed results will be presented in the following.

\section{COMPUTATIONAL DETAILS}

Our first-principles calculations are performed using the projector-augmented wave method within the density-functional theory\cite{31,32}, as implemented in the Vienna Ab-initio Simulation Package (VASP).\cite{33,34} The plane wave energy cutoff is set to 500 eV. For the exchange-correlation potential, we use the generalized gradient approximation (GGA) by Perdew, Burke, and Ernzerhof.\cite{35} The rotationally invariant GGA+U method is employed with $U=5.00$ and $J=0.64$ eV for Ti 3d.\cite{36} STO (001) layer has $\sqrt{2}\times\sqrt{2}$ periodicity with experimental lattice constant $\sqrt{2} a$ of 5.52 \AA.\cite{37} The (STO)$_4$/(YTO)$_2$ superlattice structure are fully optimized with in-plane lattice constant constrained to the STO substrare using a $\Gamma$-centered $4\times 4\times 1$ k-grid. The electronic structure calculations were performed by using a $\Gamma$-centered $12\times 12\times 1$ k-grid. Our convergence standard requires that the Hellmann-Feynmann force on each atom is less than 0.01 eV/\AA{} and the absolute total energy difference between two successive loops is smaller than $10^{-5}$ eV. The strain is defined as $\varepsilon_s = (a-a_0)/a_0\times100\%$, where $a_0$ is the lattice constant of STO, and $a$ is the in-plane lattice constant under strain. Given a strain value 4\% ($a$=5.74 \AA ), 2\% ($a$=5.63 \AA ), -2\% ($a$=5.41 \AA ), and -4\% ($a$=5.30 \AA ), the superlattice is allowed to relax sufficiently with in-plane lattice constant constrained to the given value. In addition, we change the on-site interaction $U$ into $4.00$ eV while fixing the $J$ to 0.64 eV.

\section{RESULTS AND DISCUSSION}

\begin{figure}[!htbp]
\centering  
\includegraphics[clip, width=8.5cm]{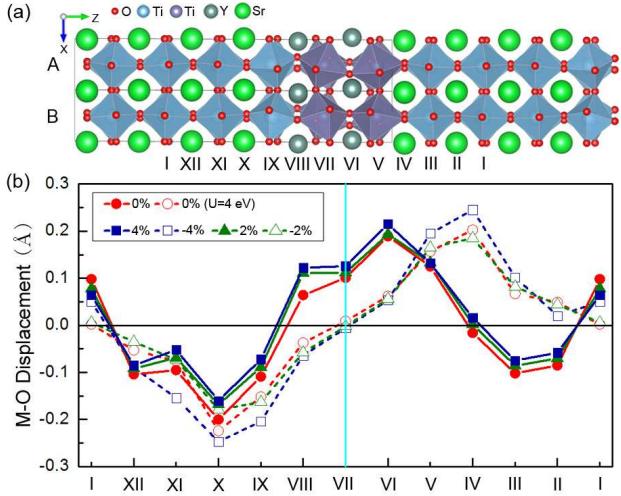}
\caption{(Color online) (a) Side view of optimized atomic structure of (STO)$_4$/(YTO)$_2$ superlattice. A and B represent two inequivalent in-plane Ti atoms in each of the TiO$_2$ monolayer. (b) The average displacements along the z axis of cation M (M=Sr, Ti, and Y) with respect to anion O in each monolayer for five strain values (0.0\%, $\pm 2.0$\%, and $\pm 4.0$\%) and $U=5$ (hidden) or $4$ (shown) eV. The negative and positive values imply that the M cations move leftward or rightward with respect to the O anions, respectively.}\label{edge}
\end{figure}

In the YTi$^{3+}$O$_3$ (001) direction, the Y$^{3+}$O$^{2-}$ and Ti$^{3+}$O$_2^{4-}$ monolayers appear alternately, carrying formal charges of +1 and -1, respectively, but in the non-polar SrTi$^{4+}$O$_3$ (001) direction both Sr$^{2+}$O$^{2-}$ and Ti$^{4+}$O$_2^{4-}$ monolayers are charge neutral. The optimized atomic structure of the (STO)$_4$/(YTO)$_2$ superlattice without strain is shown in Fig. 1 (a). Figure 1 (b) shows the calculated average displacements along the z axis of cation M (M = Sr, Ti, and Y) with respect to anion O in each monolayer of the (STO)$_4$/(YTO)$_2$ superlattice for the five strain $\varepsilon_s$ values and $U=5$ or 4 eV. Under tensile strain of 2\% and 4\%, layer-resolved change trend of M-O (M=Sr, Ti, and Y) displacement is consistent with that of the unstrained system. The negative and positive displacements per monolayer in the STO and YTO layer, respectively, providing screening to the diverging electrostatic potential in YTO layer. However, when the compressive strain $\varepsilon_s$ of -2\% or -4\% is applied, the absolute value of M-O (M=Sr, Ti, and Y) displacement for the layers labelled from II to VI is almost equivalent to those of the layers labelled from XII to VIII, indicating that there exists an inversion symmetry and the inversion center lies in the plane of TiO$_2$ layer labelled with VII. The space group for this two structural phases of YTO/STO superlattice correspond to \emph{PC} and \emph{P}21/\emph{C}, respectively.

\begin{figure}[!htbp]
\centering  
\includegraphics[clip, width=8.5cm]{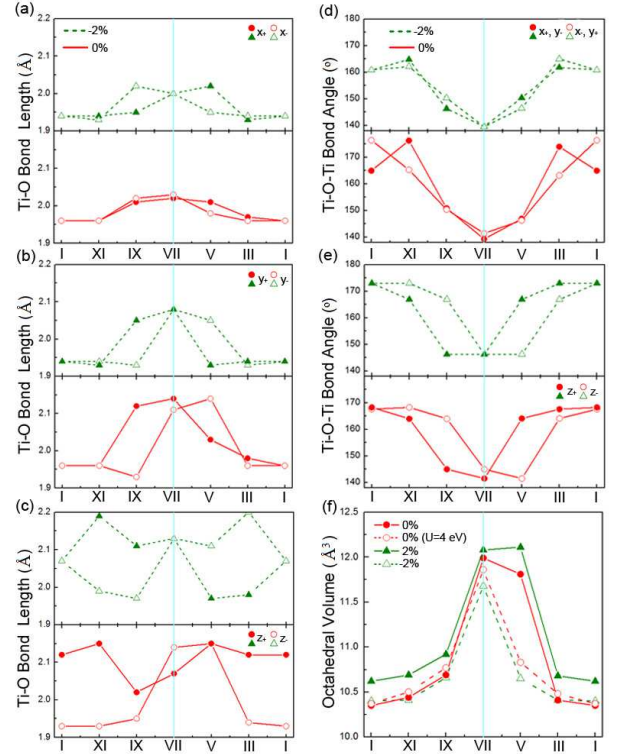}
\caption{(Color online) The optimized Ti-O bond lengths (a-c) and Ti-O-Ti bond angles (d,e) of the TiO$_{6}$ octahedra in (STO)$_4$/(YTO)$_2$ superlattice at $\varepsilon_s$=-2\% and 0\%. The six O atoms around the each Ti atom are labeled by the direction from Ti to O atoms expressed by x$_\pm$, y$_\pm$, and z$_\pm$. (f) Octahedral volumes of TiO$_{6}$ octahedra in superlattices under $\varepsilon_s$=$\pm$2\% and 0\% with $U=5$ (hidden) and/or $4$ eV.}\label{edge}
\end{figure}

In order to show more structural features of the (STO)$_4$/(YTO)$_2$ superlattices with space group of \emph{PC} and \emph{P}21/\emph{C}, we present the Ti-O bond lengths, Ti-O-Ti bond angles, and octahedral volumes of TiO$_{6}$ octahedra in the A sublattice in Fig. 2. Interchanging x$_\pm$ and y$_\pm$ in Fig. 2 gives the Ti-O bond lengths and Ti-O-Ti bond angles of TiO$_{6}$ octahedra in the B sublattice. To show the structural phase transition more clearly, we present the calculated results for $\varepsilon_s$= 0\% and -2\% only, corresponding to \emph{PC} and \emph{P}21/\emph{C}, respectively. In Fig. 2 (a)-(e), it can be clearly seen that, at $\varepsilon_s$= -2\%, the Ti-O bond lengths and Ti-O-Ti bond angles of TiO$_{6}$ octahedra in the TiO$_2$ monolayer labelled with VII along the x$_+$, y$_+$, and z$_+$ directions are exactly equal to those along the x$_-$, y$_-$, and z$_-$ directions, respectively, taking the Ti atom in the VII monolayer as the inversion center. Interchanging (x$_+$, y$_+$, z$_+$) and (x$_-$, y$_-$, z$_-$) in the monolayers III and V, we can obtain the Ti-O bond lengths and Ti-O-Ti bond angles in the monolayers IX and XI, respectively. The octahedral volume of YTO and STO in bulk with experimental lattice constant is 11.33 and 9.92 \AA$^3$, respectively. As is shown in Fig. 2 (f), compared with the TiO$_{6}$ octahedral volume of 11.99 \AA$^3$ of the unstrained system in the VII monolayer, it decreases by 4.5\% (11.45 \AA$^3$) at $\varepsilon_s$= -2\%. However, compared with the TiO$_{6}$ octahedral volume of 11.81 \AA$^3$ in the V monolayer of the unstrained system, it substantially decreases by 9.8\% (10.65 \AA$^3$) at $\varepsilon_s$= -2\%. Therefore, the TiO$_{6}$ octahedral volumes in the monolayers III and V are equal to those in the monolayers XI and IX. These properties of structural parameters in the superlattice at $\varepsilon_s=-2\%$ reflect the additional inversion symmetry.


To show the evolution of the electronic structure resulting from the structural phase transition from \emph{PC} to \emph{P}21/\emph{C}, we present in Figs. 3 (a)-(c) the spin-resolved band structures of the (STO)$_4$/(YTO)$_2$ superlattices corresponding to $\varepsilon_s$=0\%, 2\%, and -2\%. Figure 3 (a) and (b) show that the (STO)$_4$/(YTO)$_2$ superlattice is an insulator at $\varepsilon_s$= 0\% and 2\%. From Fig. 3(a)-(c), we note that the red filled bands of Ti atoms in the VII monolayer always lie below the Fermi level at $\varepsilon_s$=0\%, 2\% and -2\%, and all these Ti atoms have the magnetic moment value of 0.88 $\mu_B$, indicating that the Ti cation in this TiO$_2$ layer almost completely confines an electron with Ti$^{3+}$ and thus keeps the Mott-insulator character. Comparing Fig. 3 (c) with Fig. 3 (a), it can be seen that the d level in the V monolayer and the Fermi level move upward in the case of $\varepsilon_s$=-2\% and the electronic reconstruction takes place in the superlattice between $\varepsilon_s$=0 and $\varepsilon_s$= -2\%. The phase with $\varepsilon_s$= -2\% shows a metallic property. This implies that there exists a phase transition between $\varepsilon_s$=0 and $\varepsilon_s$=-2\%. It is also interesting that such a phase transition can take place by decreasing the parameter $U$ from 5 to 4 eV, as shown in Fig. 3(d).

\begin{figure}[!tbp]
\centering  
\includegraphics[clip, width=8.5cm]{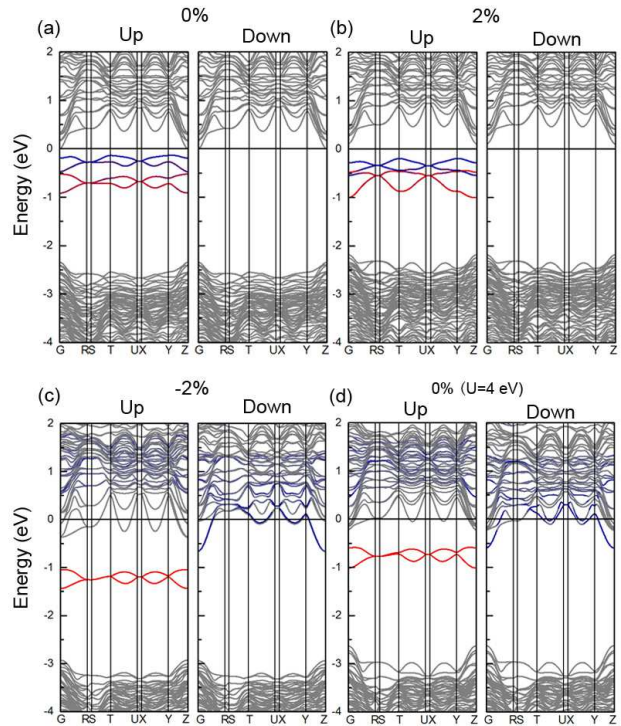}
\caption{(Color online) Spin-resolved band structures of the (STO)$_4$/(YTO)$_2$ superlattice with $U=5$ eV at $\varepsilon_s$=0\% (a), 2\% (b), and -2\% (c); and $U$= 4.00 eV and $\varepsilon_s$=0\% (d). The red and blue lines indicate the bands originated from Ti atoms in the TiO$_2$ monolayer labelled with VII and V. }\label{edge}
\end{figure}

\begin{figure}[!htbp]
\centering  
\includegraphics[clip, width=8.5cm]{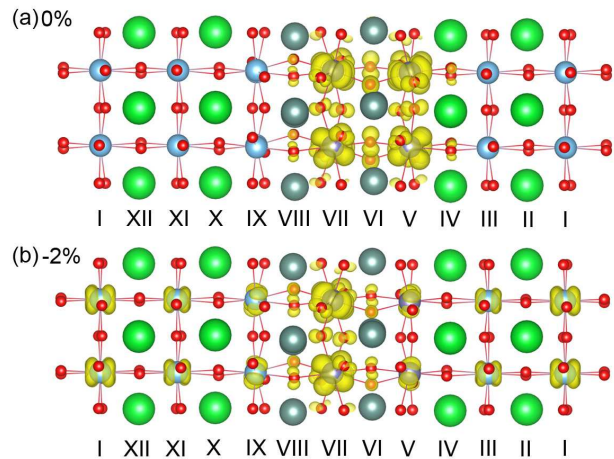}
\caption{(Color online) The yellow isosurfaces (0.004$|e|$/\AA$^3$) of the charge density from the energy window of -2 to 0 eV in the (STO)$_4$/(YTO)$_2$ superlattice at $\varepsilon_s$= 0\% (a) and -2\% (b).}\label{edge}
\end{figure}

To show the change of the electron density distribution over the interfacial electron reconstruction and the insulator-metal transition, we present in Fig. 4 (a) and (b) the charge density distribution of the (STO)$_4$/(YTO)$_2$ superlattices at $\varepsilon_s$=0\% and -2\% in the energy window between -2 and 0 eV. The charge density distributions in the V and VII monolayers shown in Fig. 4 (a) lies 0.2-0.5 and 0.5-0.9 eV below the Fermi level, respectively, corresponding to Fig. 3 (a). The electron density distribution of the electronic states near the Fermi level shown in Fig. 3(c) can be seen clearly in Fig. 4 (b). Comparing the electron density distributions between the two sides of the phase transition, it can be clearly seen that there is a large transfer of electron from the V monolayer to the other TiO$_2$ monolayers when the strain changes from 0 to -2\%. These Ti atoms exhibit mixed valence between +3 and +4, due to the electron reconstruction in the superlattices.


\begin{figure}[!htbp]
\centering  
\includegraphics[clip, width=8.5cm]{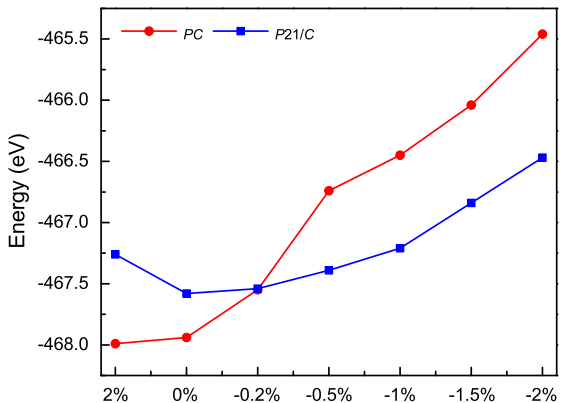}
\caption{(Color online) The total energy comparison of the A (red, \emph{PC}) and B (blue, \emph{P}21/\emph{C}) structures with the strain changing from $\varepsilon_s$= 2\% to -2\%.}\label{edge}
\end{figure}

To determine the structural phase transition point of the compressive strain, we calculate the total energies of the two structural phases as functions of the stain (from 2\% to -2\%). The calculated results are presented in Fig. 5. It is clear that the structural phase transition takes place near -0.2\%. In the YTO layer, the middle TiO$_2$ monolayer is between the two Y-O monolayers (VI and VIII), but each of the interfacial TiO$_2$ monolayers (V and IX) is inserted between one Sr-O monolayer (IV or X) and one Y-O monolayer (VI or VIII). Other TiO$_2$ monolayers each have two Sr-O monolayers of STO as nearest neighbors. We find the absence of filled $d$ energy levels of Ti atoms in the V monolayer of the YTO layer at the structural phase transition point.

It is widely recognized that the tilting of the TiO$_{6}$ octahedra controls the electron transfers between the neighbouring Ti t$_{2g}$ orbitals mediated by the O 2$p$ orbitals, thereby reducing the t$_{2g}$ bandwidth with decreasing bond angle. The variations of the Mott-Hubbard gap and electron correlations in the insulating titanates with GdFeO$_3$-type distortion have been experimentally studied by controlling the bandwidth through this distortion.\cite{38,39,40}
From Fig. 2 (d) and (e), it can be seen that the in-plane Ti-O-Ti bond angles of the TiO$_{6}$ octahedron in the V monolayer increase by about 4.5\% compared with those in the VII monolayer. Compared with that of 145.0$^\circ$ along the z$_-$ direction (across the VII, VIII, and IX monolayers) in the VII monolayer, the Ti-O-Ti bond angles of 164.1$^\circ$ is substantially larger by 13.2\% along the z$_+$ direction (across the V, IV, and III monolayers) in the V monolayer. The value along the z$_-$ direction in the V monolayer is the same as that along the z$_+$ in the VII monolayer. Thus, compared with those in the VII monolayer, these bond angles have smaller departure from 180$^\circ$ in the V monolayer of the unstrained system. Therefore, the energy level of the blue band from Ti atoms in the V monolayer is higher than that in the VII monolayer (see Fig.3). With the increasing compressive strain, the Ti-O-Ti bond angle along z$_+$ in the V monolayer continues to increase and then the original electrons in the V monolayer transfers to other TiO$_2$ monolayers due to the decreasing electron correlation. Thus, the electron reconstruction takes place in such superlattices, accompanied with structural phase transition.

Analogously, when the $U$ of Ti atoms is changed from 5 to 4.00 eV, the superlattices also becomes conductive due to the electron reconstruction in such superlattices. $U$ is the on-site Coulomb repulsion energy and reducing $U$ of Ti atom means that the strength of the electron correlation becomes weak. This will also induce the electrons in the V monolayer to transfer to other TiO$_2$ monolayers duo to the smaller departure from 180$^\circ$ of the Ti-O-Ti bond angles in the V monolayer than those in the VII monolayer.

\section{CONCLUSION}

In summary, we have found an insulator-metal phase transition of the (STO)$_4$/(YTO)$_2$ superlattices when applying compressive strain or reducing the electronic $U$ parameter. Our study shows that the metallic phases occur when the compressive strain is stronger than 0.2\% or the on-site interaction \emph{U} changes from $U=5.00$ to $U=4.00$ eV. The structure analyses reveal that, with the insulator-metal transition, there exists a structural phase transition from \emph{PC} to \emph{P}21/\emph{C} phase due to the restoring of inversion symmetry. The smaller departure from 180$^\circ$ of the Ti-O-Ti bond angles in the interfacial TiO$_2$ monolayer than those in the middle TiO$_2$ monolayer in the YTO layer of (STO)$_4$/(YTO)$_2$ superlattice make the electron correlation of Ti atoms in the V monolayer become weaker. With increasing compressive strain or decreasing $U$, the electron correlation of Ti atoms in the interfacial TiO$_2$ monolayer continues to become weak, and then the $d$ energy level of Ti atoms in this monolayer moves upward, inducing the electron reconstruction in such superlattices. It is believed that these findings can improve our understanding of such oxide superlattices.

\section{ACKNOWLEDGMENTS}
This work is supported by the Nature Science Foundation of China (Grant No. 11174359 and No. 11574366), by the Department of Science and Technology of China (Grant No. 2016YFA0300701 and No. 2012CB932302), and by the Strategic Priority Research Program of the Chinese Academy of Sciences (Grant No.XDB07000000). The calculations were performed in the Milky Way \#2 supercomputer system at the National Supercomputer Center of Guangzhou.

\end{document}